\title{NBA PCA Manuscript - JSA Version}

\documentclass[preprint]{imsart}

\RequirePackage[OT1]{fontenc}
\RequirePackage{amsthm,amsmath,amssymb}
\RequirePackage[numbers]{natbib}
\RequirePackage[colorlinks,citecolor=blue,urlcolor=blue]{hyperref}

\usepackage{lscape}
\usepackage{hyperref}
\usepackage[margin=1.2in]{geometry}
\usepackage[english]{babel}
\usepackage[utf8]{inputenc}
\usepackage{mathrsfs}
\usepackage{graphicx}
\usepackage{hyperref}
\usepackage{listings}
\usepackage{caption,color}
\usepackage{subcaption}
\usepackage{enumitem}
\usepackage{mathtools}
\usepackage{scalerel}
\usepackage{mathtools}
\usepackage[pdftex,dvipsnames]{xcolor}  
\usepackage{xargs}
\usepackage[colorinlistoftodos,prependcaption,textsize=tiny]{todonotes}
\usepackage{multirow}

\date{\today}

\begin{document}

\begin{frontmatter}
\title{A Scalable Framework for NBA Player and Team Comparisons Using Player Tracking Data}
\runtitle{A Scalable Framework for NBA Player and Team Comparisons Using Player Tracking Data}

\begin{aug}

\author{\fnms{Scott} \snm{Bruce}\ead[label=e1]{scott.bruce@temple.edu}}

\runauthor{Bruce 2015}

\affiliation{Department of Statistics, Temple University}

\address{Temple University\\
Department of Statistics\\
1801 Liacouras Walk\\
Philadelphia, Pennsylvania, 19122, U.S.A.\\
}
\end{aug}

\begin{abstract}
The release of NBA player tracking data greatly enhances the granularity and dimensionality of basketball statistics used to evaluate and compare player performance.  However, the high dimensionality of this new data source can be troublesome as it demands more computational resources and reduces the ability to easily analyze and interpret findings.  Therefore, we must find a way to reduce the dimensionality of the data set while retaining the ability to differentiate and compare player performance.  

In this paper, Principal Component Analysis (PCA) is used to identify four principal components that account for 68\% of the variation in player tracking data from the 2013-2014 regular season and intuitive interpretations of these new dimensions are developed by examining the statistics that influence them the most.  In this new high variance, low dimensional space, you can easily compare statistical profiles across any or all of the principal component dimensions to evaluate characteristics that make certain players and teams similar or unique.  A simple measure of similarity between two player or team statistical profiles based on the four principal component scores is also constructed.  The Statistical Diversity Index (SDI) allows for quick and intuitive comparisons using the entirety of the player tracking data. As new statistics emerge, this framework is scalable as it can incorporate existing and new data sources by reconstructing the principal component dimensions and SDI for improved comparisons.

Using principal component scores and SDI, several use cases are presented for improved personnel management.  Team principal component scores are used to quickly  profile and evaluate team performances across the NBA and specifically to understand how New York's lack of ball movement negatively impacted success despite high average scoring efficiency as a team.  SDI is used to identify players across the NBA with the most similar statistical performances to specific players.  All-Star Tony Parker and shooting specialist Anthony Morrow are used as two examples and presented with in-depth comparisons to similar players using principal component scores and player tracking statistics.  This approach can be used in salary negotiations, free agency acquisitions and trades, role player replacement, and more. 

\textbf{Keywords:}  Principal component analysis, NBA player tracking data, statistical diversity index, dimension reduction, personnel management, National Basketball Association

\end{abstract}
\end{frontmatter}

\section{Introduction}
The National Basketball Association (NBA) launched a public database in September 2013 containing over 80 new statistics captured by STATS LLC through their innovative sportVU player tracking camera systems\cite{tracking}.  The cameras capture and record the location of players on the court as well as the location of the ball, and the data are used to derive many different interesting and useful stats that expand greatly upon the traditional stats available for analysis of basketball performance.  We can now break down shot attempts and points by shot selection (e.g. driving shots, catch and shoot shots, pull up shots), assess rebounding ability for contested and uncontested boards, and even look at completely new statistics like average speed and distance and opponent field goal percentage at the rim.  The availability of such data enables fans and analysts to dig into the data and uncover insights previously not possible due to the limited nature of the data at hand. \\

For example, techniques to uncover different 'positions' based on grouping statistical profiles have become increasingly popular.  It reflects the mindset of current NBA coaches and general managers who are very much aware of the different types of players beyond the five traditional roles, but a recent proposal\cite{5to13} has received criticism for its unintuitive groupings and inability to separate out the impact of player talent\cite{slate}. The NBA player tracking data has the ability to differentiate player performance across more dimensions than before (e.g. shot selection, possession time, physical activity, etc.) which can provide better ways to evaluate the uniqueness and similarities across NBA player abilities and playing styles. Additionally, many research methods for basketball analysis rely on estimation of possessions and other stats to produce offensive and defensive ratings\cite{start}.  With the ability to track players' time of possession and proximity to players in possession of the ball through player tracking, we can develop more accurate representations of possessions and better player offensive and defensive efficiency metrics.\\

However, the high dimensionality of this new data source can be troublesome as it demands more computational resources and reduces the ability to easily analyze and interpret findings.  We must find a way to reduce the dimensionality of the data set while retaining the ability to differentiate and compare player performance.  One method that is particularly well-suited for this application is Principal Component Analysis (PCA) which identifies the dimensions of the data containing the maximum variance in the data set.  This article applies PCA to the NBA player tracking data to discover four principal components that account for 68\% of the variability in the data for the 2013-2014 regular season.  These components are explored in detail by examining the player tracking statistics that influence them the most and where players and teams fall along these new dimensions.  \\

In addition to exploring player and team performances through the principal components, a simple measure of similarity in statistical profiles between players and teams based on the principal components is proposed.  The Statistical Diversity Index (SDI) can be calculated for any pairwise player combination and provides a fast and intuitive method for finding players with similar statistical performances along any or all of the principal component dimensions.  This approach is also advantageous from the standpoint of scalability. The possibilities to derive new statistics from the player tracking data are endless, so as new statistics emerge, this approach can again be applied using the new and existing data to reconstruct the principal components and SDI for improved player evaluation and comparisons.  
\\

Numerous applications in personnel management exist for the use of SDI and the principal component scores in evaluating and comparing player and team statistical performances.  Two specific case studies are presented to show how these tools can be used to quickly identify players with similar statistical profiles to a certain player of interest for the purpose of identifying less expensive, similarly skilled players or finding suitable replacement options for a key role player within the organization.  \\   

This article is organized as follows.  Section \ref{sec:data} describes the player tracking data and data processing.  Section \ref{sec:PCA} provides the analysis and interpretations for the four principal components in detail, showing how players and teams can be compared across these new dimensions.  Section \ref{sec:SDI} introduces the calculation for SDI and two case studies where principal component scores and SDI are used to find players with similar statistical profiles to All-Star Tony Parker and role player Anthony Morrow for personnel management purposes.  Section \ref{sec:conclusion} concludes the article with final remarks.

\section{NBA Player Tracking Data}
\label{sec:data}
\subsection{Data Description}
Currently, there are over 90 new player tracking statistics, and data for all 482 NBA players from the 2013-2014 regular season are available.  Separate records exist for players who played for different teams throughout the season, including a record for overall performance across all teams.  Brief descriptions of the newly available statistics adapted from the NBA player tracking statistics website\cite{statsdesc} are provided for reference and will be helpful in better understanding the analysis going forward.  \\

\textbf{Shooting}\\
Traditional shooting statistics are now available for different shot types:
\begin{itemize}
	\item \emph{Pull up shots} - shots taken 10 feet away from the basket where player takes 1 or more dribbles prior to shooting
	\item \emph{Driving shots} - shots taken where player starts 20 or more feet away from the basket and dribbles less than 10 feet away from the basket prior to shooting
	\item \emph{Catch and shoot shots} - shots taken at least 10 feet away from the basket where player possessed the ball for less than 2 seconds and took no dribbles prior to shooting
\end{itemize}

\textbf{Assists}\\
New assist categories are available that enhance understanding of offensive contribution:
\begin{itemize}
	\item \emph{Assist opportunities} - passes by a player to another player who attempts a shot and if made would be an assist
	\item \emph{Secondary assists} - passes by a player to another player who receives an assist
	\item \emph{Free throw assists} - passes by a player to another player who was fouled, missed the shot if shooting, and made at least one free throw
	\item \emph{Points created by assists} - points created by a player through his assists
\end{itemize}

\textbf{Touches}\\
Location of possessions provides insight into style of play and scoring efficiency:
\begin{itemize}
	\item \emph{Front court touches} - touches on his team's offensive half of the court
	\item \emph{Close touches}  - touches that originate within 12 feet of the basket excluding drives
	\item \emph{Elbow touches} - touches that originate within 5 feet of the edge of the lane and the free throw line inside the 3-point line
	\item \emph{Points per touch} - points scored by a player per touch
\end{itemize}

\textbf{Rebounding}\\
New rebounding statistics incorporate location and proximity of opponents:
\begin{itemize}
	\item \emph{Contested rebound} - rebounds where an opponent is within 3.5 feet of the rebound
	\item \emph{Rebounding opportunity} - when a player is within 3.5 feet of a rebound
	\item \emph{Rebounding percentage} - rebounds over rebounding opportunities
\end{itemize}

\textbf{Rim Protection}\\
When a player is within 5 feet of the basket and within 5 feet of the offensive shooter, opponents' shooting statistics are available to measure how well a player can protect the basket.

\vspace{0.25cm}

\textbf{Speed and Distance}\\
Players' average speeds and distances traveled per game are also captured and broken out by offensive and defensive plays.

\subsection{Data Processing}
Only players who played at least half the 2013-2014 regular season, 41 games, are included in this analysis.  This restriction is made to reduce the influence of player statistics derived from only a few games played.  Also fields containing season total statistics and per game statistics are dropped from the analysis since they could be influenced by number of games and minutes played throughout the season.  Instead, per 48 minutes, per touch, and per shot statistics are used.  The final data set contains 360 player records each containing 66 different player tracking statistics. \\


\section{Principal Component Analysis}
\label{sec:PCA}
With numerous player tracking statistics already available and the potential to develop infinitely many more, it is increasingly difficult to extract meaningful and intuitive insights on player comparisons.  Now that the data are available for more granular and detailed comparisons, a methodology is needed that can analyze the entirety of the data set to extract a handful of dimensions for comparisons.  These dimensions should be constructed in a way that ensures optimality in differentiating players (i.e. dimensions should retain the maximum amount of player separability possible from the original data) and can be understood in terms of the original statistics.  \\

Principal Component Analysis (PCA), developed by Karl Pearson in 1901\cite{Pearson} and later by Hotelling in 1933\cite{Hotelling}, is a particularly well-suited statistical tool that can accomplish this task through identifying uncorrelated linear combinations of player tracking statistics that contain maximum variance.  Interested readers can find a brief technical introduction to PCA in Appendix \ref{app:pcaintro}.   Components of high variance help us to better differentiate player performance in these directions in hopes that the majority of the variance will be contained in a small subset of components.  This simple and intuitive approach to dimension reduction provides a platform for player comparisons across dimensions that best separate players by statistical performance and can be implemented without expensive proprietary solutions, providing more visibility into how the method works at little to no additional cost. 

\subsection{Dimension Reduction}
PCA is sensitive to different variable scalings in the original data set such that variables with larger variances may dominate the principal components if not adjusted.  To set every statistic on equal footing, all statistics are standardized with mean 0 and variance 1 prior to conducting the analysis.\\

\begin{figure}
\centering
\includegraphics[width=0.5\textwidth]{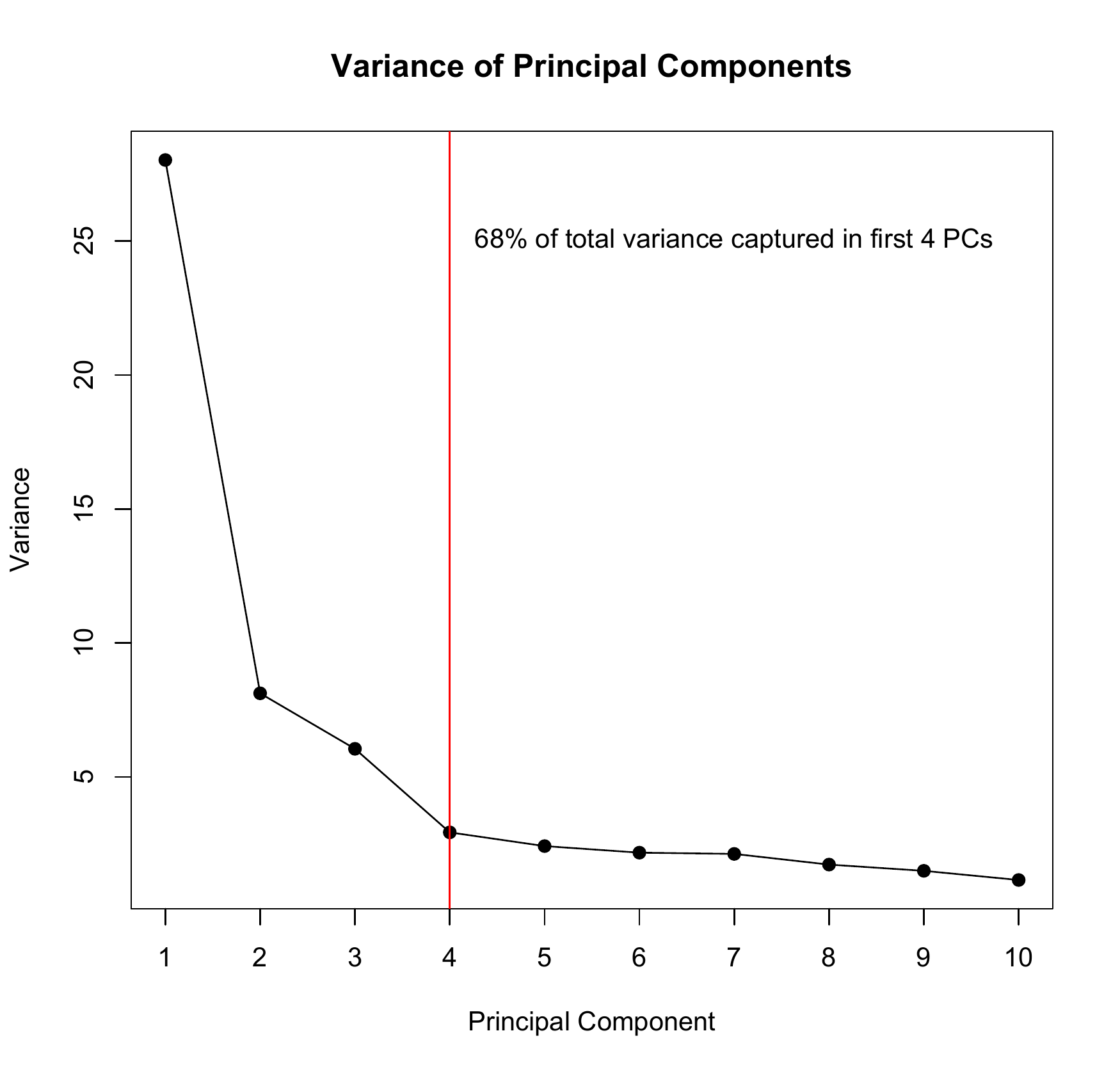}
\caption{\label{fig:scree} Variance captured by first ten principal components (color online).}
\end{figure}

PCA is most useful when the majority of the total variance across variables are captured by only a few of the principal components, thus the dimension reduction.  If this is the case for the NBA player tracking data set, it means that we are able to retain the ability to differentiate player performance without having to operate in such a high dimensional space.  Figure \ref{fig:scree} shows the variance captured by each principal component.  Note the variance captured in the first principal component is very high and decreases drastically through the first four components.  After that, the change in variance is relatively flat, forming an elbow shape in the plot.  This means that the variances captured by the fifth component onward are very similar and much smaller than the first four components.  Moreover, the first four components capture 68\% of the variance across the original variables, so we utilize these four components going forward to analyze and compare player performance and playing styles.

\subsection{Principal Components}
\begin{figure}
\centering
\begin{subfigure}{.5\textwidth}
  \centering
  \includegraphics[width=1\linewidth]{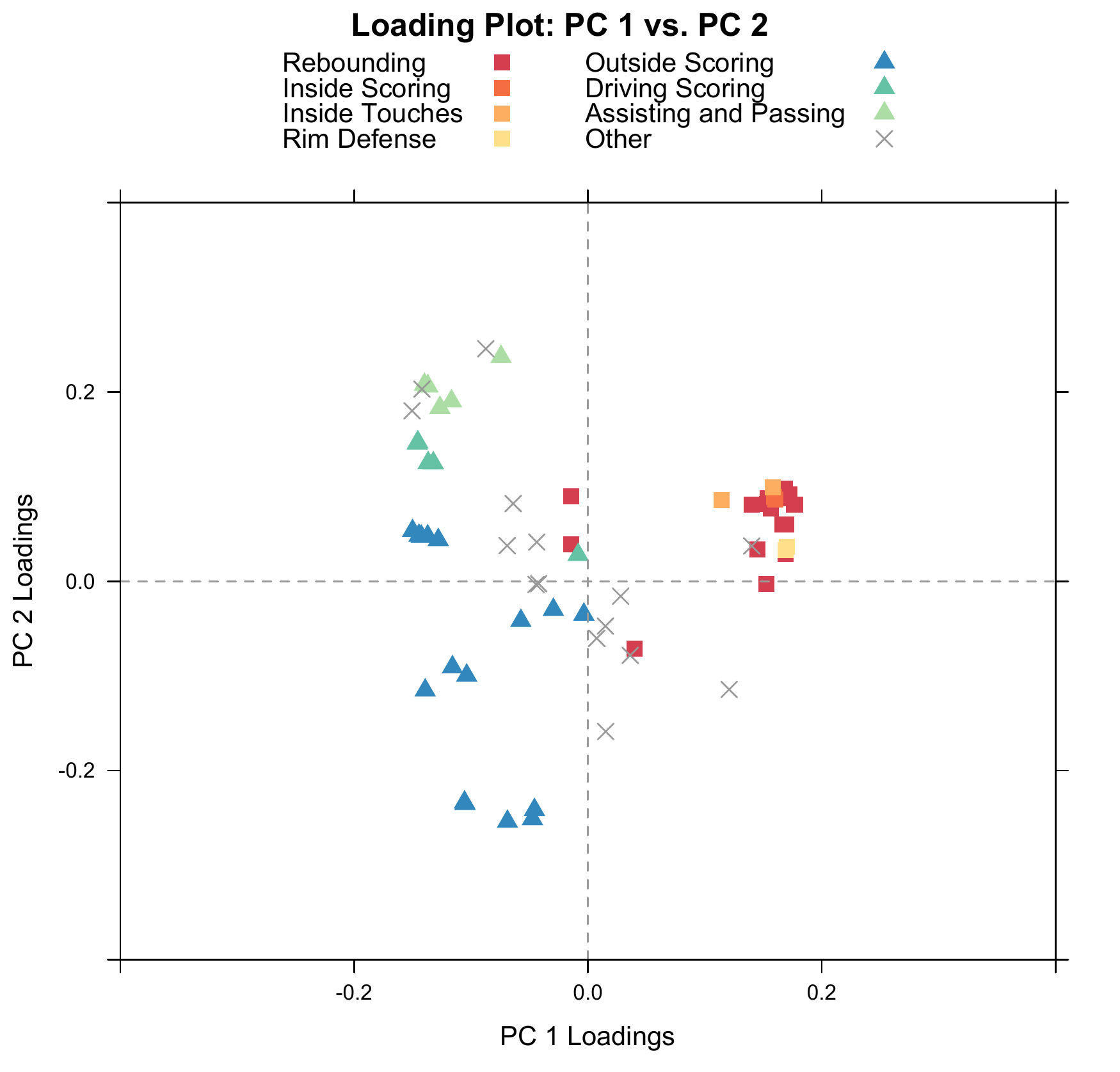}
\end{subfigure}%
\begin{subfigure}{.5\textwidth}
  \centering
  \includegraphics[width=1\linewidth]{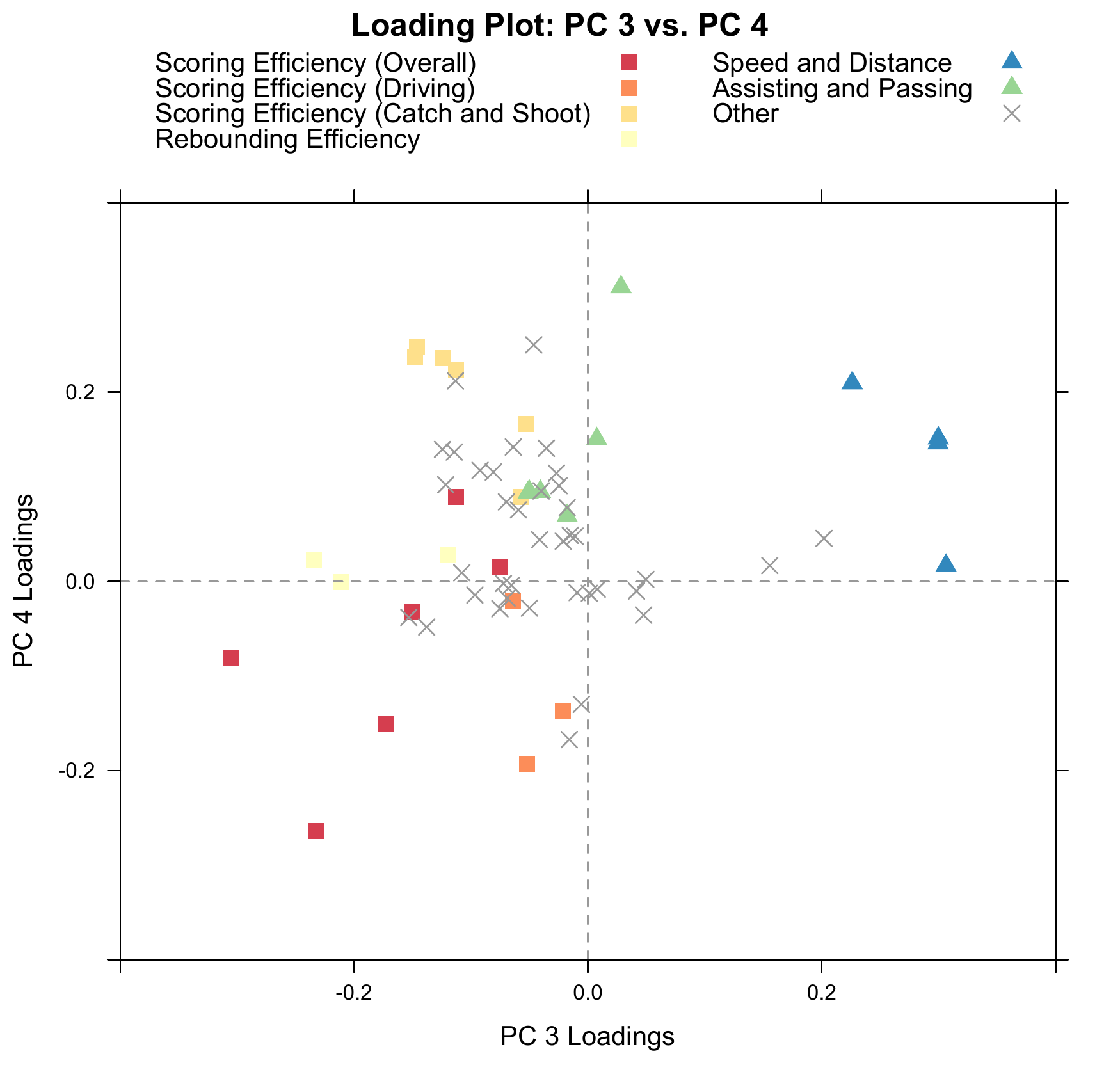}
\end{subfigure}
\centering
\caption{Categorized loading coefficients for all statistics (color online).}
\label{fig:pcload}
\end{figure}

Each principal component (PC) is a linear combination of the original variables in the dataset.  There is a vector for each principal component containing the coefficients associated with each of the variables in the original data set and are called \emph{loading vectors}. These describe the influence of each variable for each principal component and are used to interpret these new dimensions in terms of the original variables.  Figure \ref{fig:pcload} plots the categorized loading coefficients for the four principal components and is explored in detail in the following sections.    Variables can have a positive or negative contribution to the principal component.  While the sign is arbitrary, understanding which variables contribute positively or negatively can help with interpreting the principal components. The most important statistics for each component are presented in the following sections, but tables containing all loading coefficients for variables contributing significantly to the principal components are also available in Appendix \ref{app:loadings} for more details. \\

Each player can then be given a set of PC scores by multiplying the standardized statistics by their corresponding loading coefficients and then taking a sum (see Appendix \ref{app:pcaintro} for details).  Figure \ref{fig:pcscore} contains plots of the PC scores for all players with a select few noted for illustration.  Using the loading and score plots here, we can begin to understand and interpret what these new dimensions are capturing and use them for player comparisons.

\begin{figure}
\centering
\begin{subfigure}{.5\textwidth}
  \centering
  \includegraphics[width=1\linewidth]{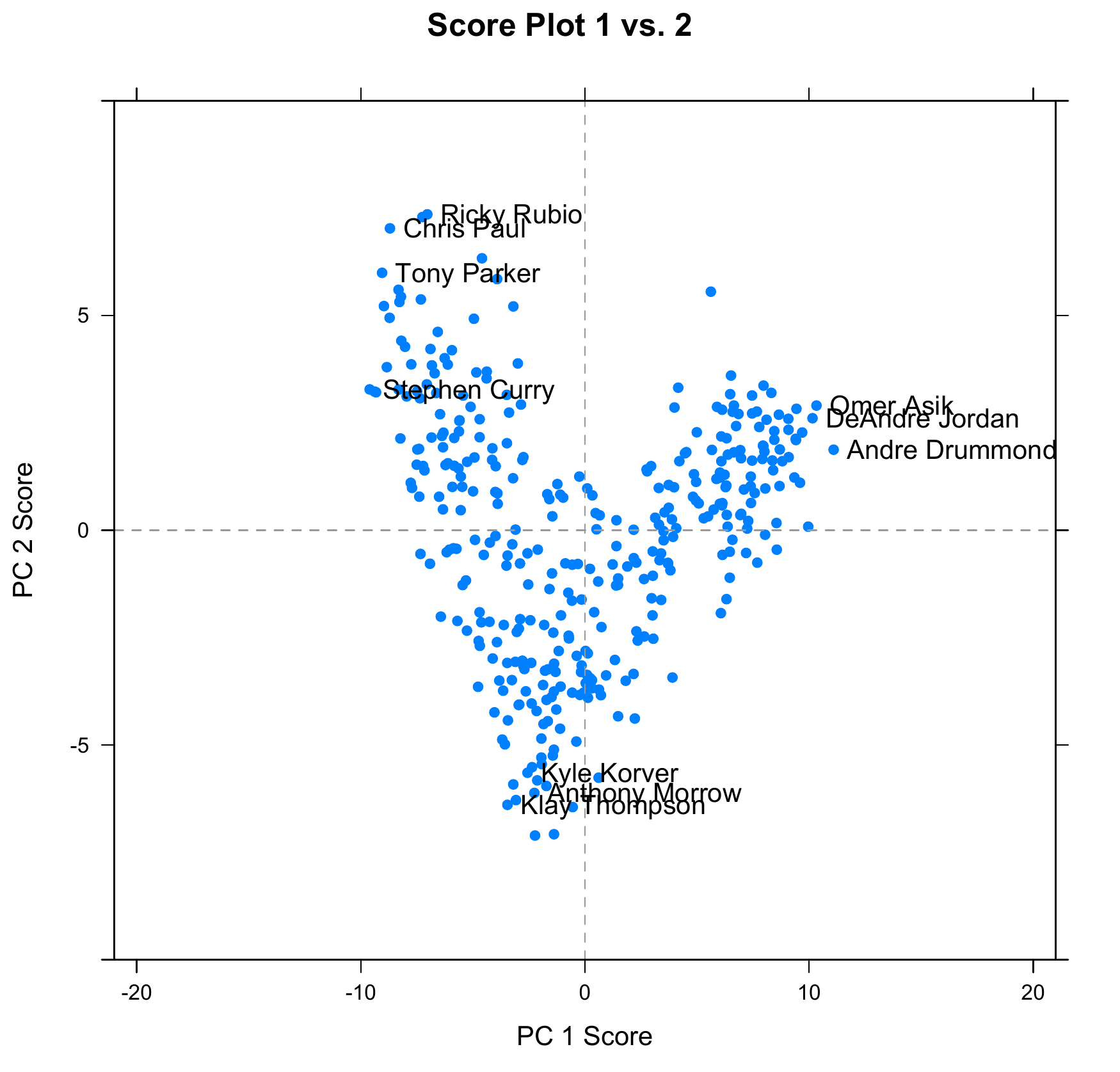}
\end{subfigure}%
\begin{subfigure}{.5\textwidth}
  \centering
  \includegraphics[width=1\linewidth]{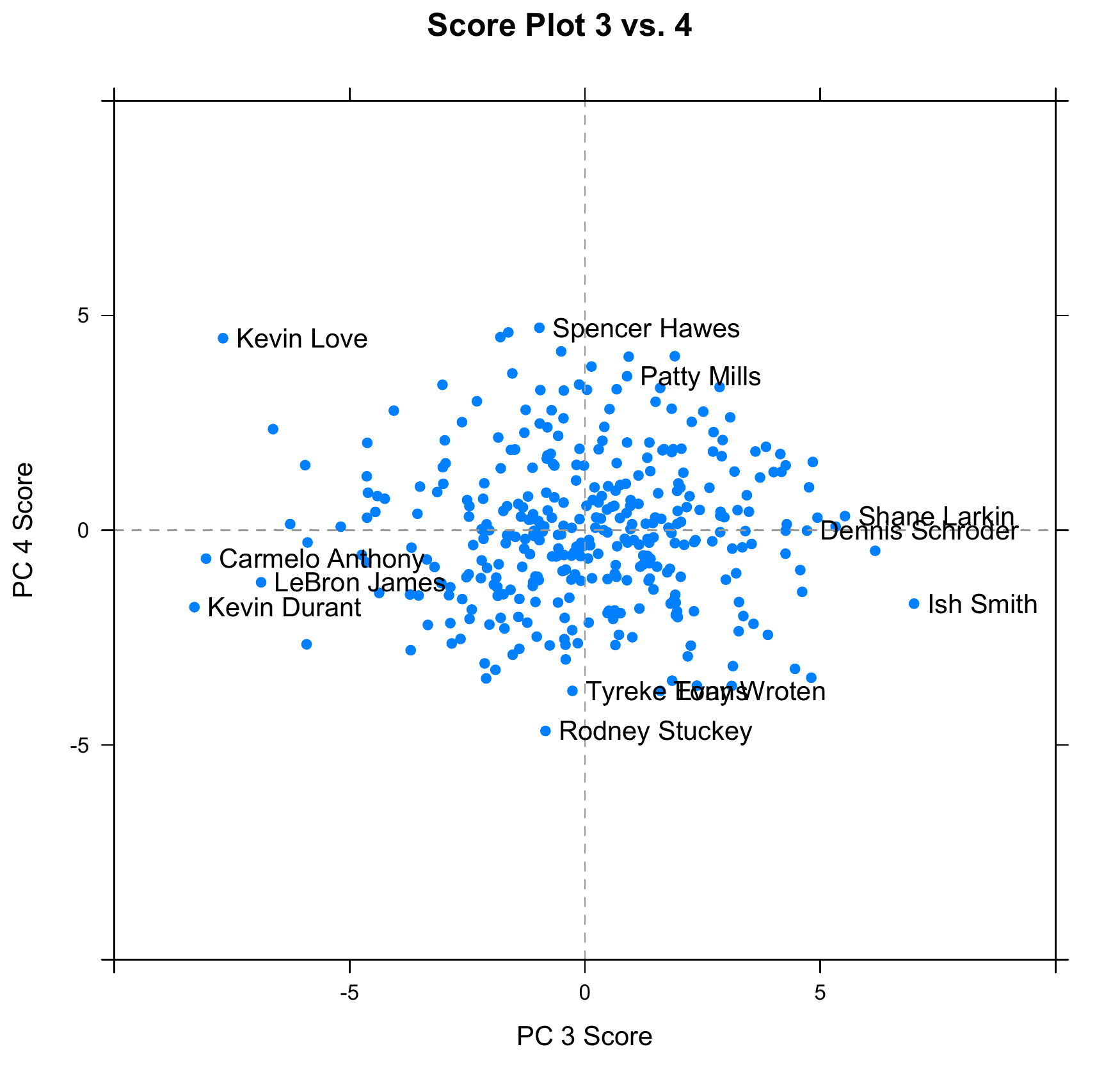}
\end{subfigure}
\centering
\caption{Player PC scores for first four components (color online).}
\label{fig:pcscore}
\end{figure}

\subsubsection{PC 1: Inside vs. Outside}

The first principal component accounts for the most variation, 42\% of the total variance.  Table \ref{tab:pc1top} lists statistics with highly positive and negative loadings for PC 1, and refer back to Figure \ref{fig:pcload} for categorized PC 1 loadings for all statistics.  These are used to better understand the meaning of the scores along this dimension.  Players with positive scores for PC 1 are able to secure rebounds of all kinds and are responsible for defending the rim and close shots.  Notable examples are Andre Drummond, DeAndre Jordan, and Omer Asik.  While players with negative scores for PC 1 drive the basketball to the hoop and take pull up and catch and shoot shots often, which implies they tend to be outside players.  These players also tend to possess the ball more often and generate additional offense through assists.  Examples here are Stephen Curry, Tony Parker, and Chris Paul.

\begin{table}[h]

\parbox{.45\linewidth}{
\centering
\begin{tabular}{|c|l|}
\hline
Loading & Statistic \\\hline

0.177 & Contested Rebounds  \\
0.176 & Rebound Opportunities \\
0.172 & Offensive Rebounds\\
0.172 & Total Rebounds \\
0.171 & Contested Offensive Rebounds \\
0.170 & Opponent Shot Attempts at the Rim \\
0.170 & Contested Defensive Rebounds \\
0.169 & Uncontested Rebounding Efficiency \\
0.169 & Opponent Made Shots at the Rim \\
0.169 & Offensive Rebound Opportunities \\
\hline
\end{tabular}
\vspace{0.25cm}
}
\hfill
\parbox{.45\linewidth}{\begin{tabular}{|c|l|}
\hline
Loading & Statistic \\\hline
-0.150 & Front Court Touches  \\
-0.150 & Pull Up Shot Attempts  \\
-0.146 & Drives  \\
-0.146 & Team Driving Points \\
-0.145 & Pull Up Points \\
-0.143 & Pull Up Made Shots \\
-0.142 & Time of Possession \\
-0.140 & Assist Opportunities \\
-0.139 & Catch and Shoot 3-Point Shooting Efficiency \\
-0.137 & Points Created by Assists \\
\hline
\end{tabular}
\vspace{0.25cm}
}
\caption{\label{tab:pc1top}Top 10 largest positive(left) and negative(right) loadings for PC 1 (per 48 minutes unless stated otherwise).}
\end{table}

\subsubsection{PC 2: Assist and Drive vs. Catch and Shoot}

PC 2 accounts for another 12\% of the total variance and Table \ref{tab:pc2top} lists statistics with highly positive and negative loadings.  Also refer to Figure \ref{fig:pcload} for categorized PC 2 loading coefficients.  Players with positive PC 2 scores generate offense mainly through assists and driving shots.  These players tend to possess the ball often and either kick the ball to teammates for shot attempts or drive the ball to the basket.  Many point guards fall into this category with examples like Ricky Rubio, Tony Parker, and Chris Paul.  Players with negative PC 2 scores provide offense primarily through catch and shoot shots and are very efficient scorers, especially from behind the 3-point arc.  Primary examples are Klay Thompson, Kyle Korver, and Anthony Morrow.

\begin{table}[h]

\parbox{.45\linewidth}{
\centering
\begin{tabular}{|c|l|}
\hline
Loading & Statistic \\\hline

0.246 & Touches  \\
0.237 & Passes \\
0.208 & Assist Opportunities\\
0.206 & Assists \\
0.206 & Points Created by Assists \\
\hline
\end{tabular}
\vspace{0.25cm}
}
\hfill
\parbox{.45\linewidth}{\begin{tabular}{|c|l|}
\hline
Loading & Statistic \\\hline
-0.254 & Catch and Shoot Points  \\
-0.251 & Catch and Shoot Attempts  \\
-0.242 & Catch and Shoot Made Shots  \\
-0.235 & Catch and Shoot 3-Point Attempts \\
-0.233 & Catch and Shoot 3-Point Made Shots \\
\hline
\end{tabular}
\vspace{0.25cm}
}
\caption{\label{tab:pc2top}Top 5 largest positive(left) and negative(right) loadings for PC 2 (per 48 minutes unless stated otherwise).}
\end{table}

\subsubsection{PC 3: Scoring and Rebounding Efficiency vs. Speed}

Table \ref{tab:pc3top} lists statistics with highly positive and negative loadings for PC 3 which explains 9\% of the total variance.  Also refer to Figure \ref{fig:pcload} for categorized PC 3 loading coefficients.  Players with positive PC 3 scores are extremely quick on both sides of the ball and cover a lot of ground while on the court.  Some examples are Ish Smith, Shane Larkin, and Dennis Schroder.  Players with negative PC 3 scores are largely responsible for scoring when on the court and provide a significant amount of offensive production per 48 minutes.  Scoring and rebounding efficiency characterize many of the superstars in the NBA with players like Kevin Durant, Carmelo Anthony, and LeBron James touting highly negative PC 3 scores.

\begin{table}[h]

\parbox{.45\linewidth}{
\centering
\begin{tabular}{|c|l|}
\hline
Loading & Statistic \\\hline

0.306 & Average Defensive Speed  \\
0.300 & Distance \\
0.300 & Average Speed\\
0.226 & Average Offensive Speed \\
0.202 & Opponent Points at the Rim \\
\hline
\end{tabular}
\vspace{0.25cm}
}
\hfill
\parbox{.45\linewidth}{\begin{tabular}{|c|l|}
\hline
Loading & Statistic \\\hline
-0.305 & Points  \\
-0.235 & Rebounding Efficiency  \\
-0.232 & Points per Touch  \\
-0.212 & Defensive Rebounding Efficiency \\
-0.173 & Points per Half Court Touch \\
\hline
\end{tabular}
\vspace{0.25cm}
}
\caption{\label{tab:pc3top}Top 5 largest positive(left) and negative(right) loadings for PC 3 (per 48 minutes unless stated otherwise).}
\end{table}

\subsubsection{PC 4: Catch and Pass/Shoot vs. Slash}

Table \ref{tab:pc4top} lists statistics with highly positive and negative loadings for PC 4 which accounts for another 4\% of the total variance. Also refer to Figure \ref{fig:pcload} for categorized loading coefficients for PC 4.  This component is characterized by players' tendencies when they receive possession of the ball.  Players with positive PC 4 scores tend to pass or convert catch and shoot shots when the ball goes their way (e.g. Kevin Love, Spencer Hawes, and Patty Mills) while players with negative PC 4 scores tend to drive the ball and score efficiently when they get touches (e.g. Tyreke Evans, Rodney Stuckey, and Tony Wroten).  

\begin{table}[h]

\parbox{.45\linewidth}{
\centering
\begin{tabular}{|c|l|}
\hline
Loading & Statistic \\\hline

0.311 & Passes  \\
0.250 & Touches \\
0.248 & Catch and Shoot Made Shots\\
0.237 & Catch and Shoot Shooting Efficiency \\
0.236 & Catch and Shoot Points \\
\hline
\end{tabular}
\vspace{0.25cm}
}
\hfill
\parbox{.45\linewidth}{\begin{tabular}{|c|l|}
\hline
Loading & Statistic \\\hline
-0.264 & Points per Touch  \\
-0.193 & Drives  \\
-0.167 & Driving Shot Attempts  \\
-0.151 & Points per Half Court Touch \\
-0.137 & Team Driving Points \\
\hline
\end{tabular}
\vspace{0.25cm}
}
\caption{\label{tab:pc4top}Top 5 largest positive(left) and negative(right) loadings for PC 4 (per 48 minutes unless stated otherwise).}
\end{table}

\subsection{Team PC Scores}
Not only can we characterize players by principal components, but teams can also be profiled along these new dimensions as well.  There are numerous ways to aggregate the player PC scores to form a team-level score, but here a simple weighted average is used.  The $k$th team PC score, $t_{k(team)}$, can be found by taking an average of the $k$th PC scores across the $n$ players weighted by the minutes played throughout the season, $m_i, i=1,\ldots,n$.   

\begin{equation}
t_{k(team)} = \frac{\sum_{i=1}^n m_i*t_{k(i)}}{\sum_{i=1}^n m_i} \qquad k=1,\ldots,4
\end{equation}

\begin{figure}[h]
\centering
\includegraphics[width=.97\textwidth]{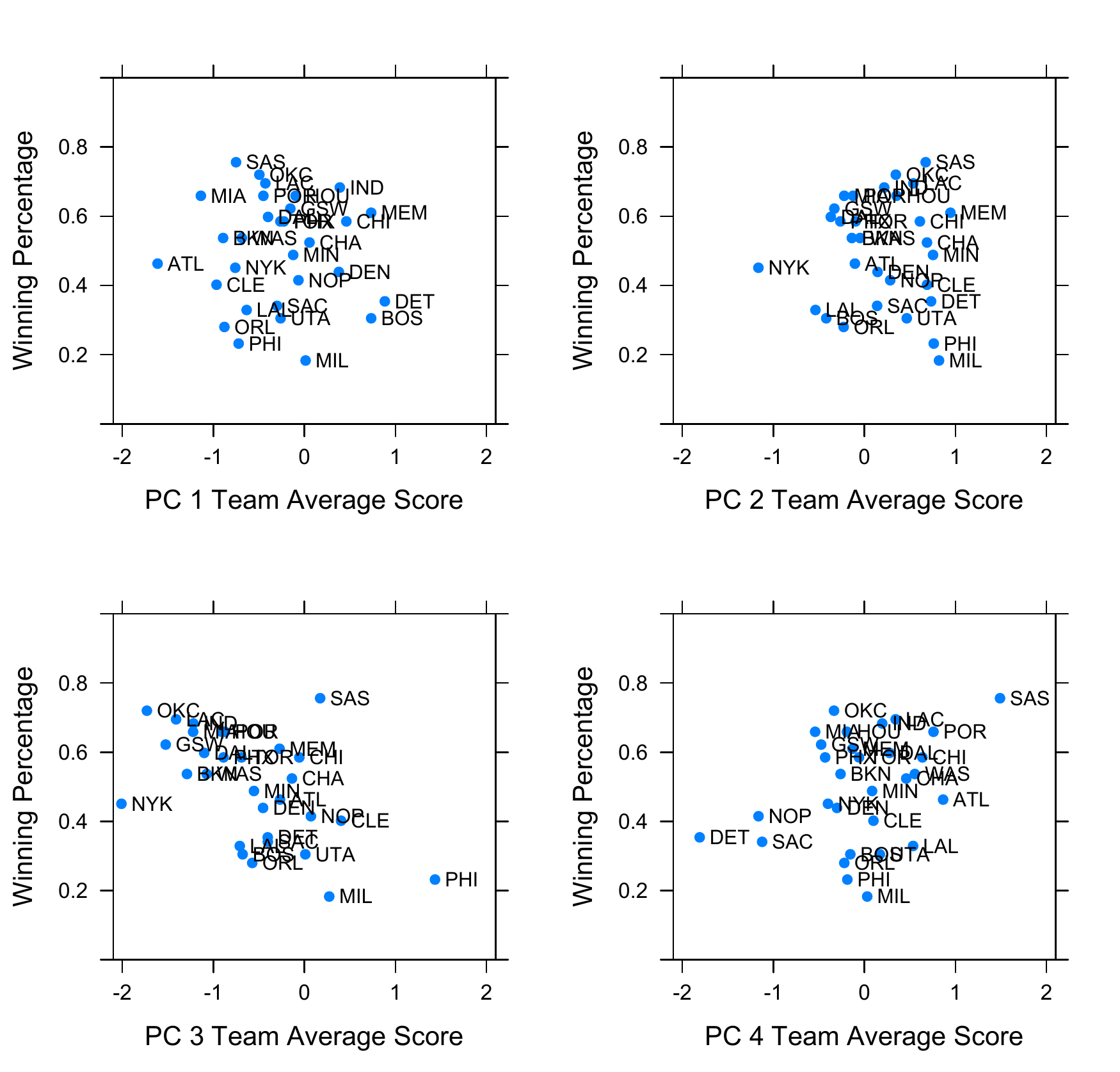}
\caption{\label{fig:pcscoreteam} Team average PC scores vs. 2013-2014 regular season winning percentage (color online).}
\end{figure}

Figure \ref{fig:pcscoreteam} shows the distribution of all NBA teams across these dimensions as well as their corresponding 2013-2014 regular season winning percentage.  This view is useful in seeing the differences and similarities in team playing styles and how they impact success.  \\\\
For example, the New York Knicks had an extremely negative PC 2 score.  Further investigation shows it is partially the result of catch and shoot offense from J.R. Smith, Andrea Bargnani, and Tim Hardaway Jr. who were all in the top 50 in catch and shoot points per 48 minutes.  However, another major factor is that 8 of the 12 New York players were below average in passes per 48 minutes (average was 58 passes per 48 minutes) which is indicative of poor ball movement.  All-Star Carmelo Anthony is in this group and has long been labeled a "ball hog"\cite{Melo} which is supported by his below average passing and above average number of touches and scoring.  In fact, Anthony's top 10 performance in points per 48 minutes, points per touch, and rebounding efficiency helped earn his team the most negative PC 3 team average score.  Note that team average PC 3 score is negatively correlated with winning percentage, yet the Knicks won only 37 games and failed to make the playoffs.  

To better understand how these team average PC scores impact winning, Table \ref{tab:reg} contains the results from a multiple linear regression analysis on winning percentage.  Note that negative PC 3 scores are highly correlated with winning while positive PC 2 and PC 4 scores are also highly correlated with winning.  Negative PC 3 scores are associated with high average scoring and rebounding efficiency.  Referring back to Tables \ref{tab:pc2top} and \ref{tab:pc4top}, passes, touches, and assists contribute positively to PC 2 and PC 4 scores.  Regarding New York's lackluster season, it seems that Anthony's great offensive production was not enough to offset the negative impact of extremely poor passing and ball movement.\\

\begin{table}[h]
\begin{tabular}{|l|l|l|l|}
\hline
Term & Coefficient & Std Error & $p$-value \\
\hline
Intercept & 0.35 & 0.04 & $<$0.001 \\
PC 1 Score & -0.01 & 0.04 & 0.758 \\
PC 2 Score & 0.17 & 0.06 & 0.005 \\
PC 3 Score & -0.20 & 0.04 & $<$0.001 \\
PC 4 Score & 0.09 & 0.03 & 0.013 \\
\hline
\end{tabular}
\vspace{0.25cm}
\caption{\label{tab:reg}Coefficient estimates in regression of team average PC scores on winning percentage ($R^2$ = 0.59).}
\end{table}

\begin{figure}[h]
\centering
\includegraphics[width=.8\textwidth]{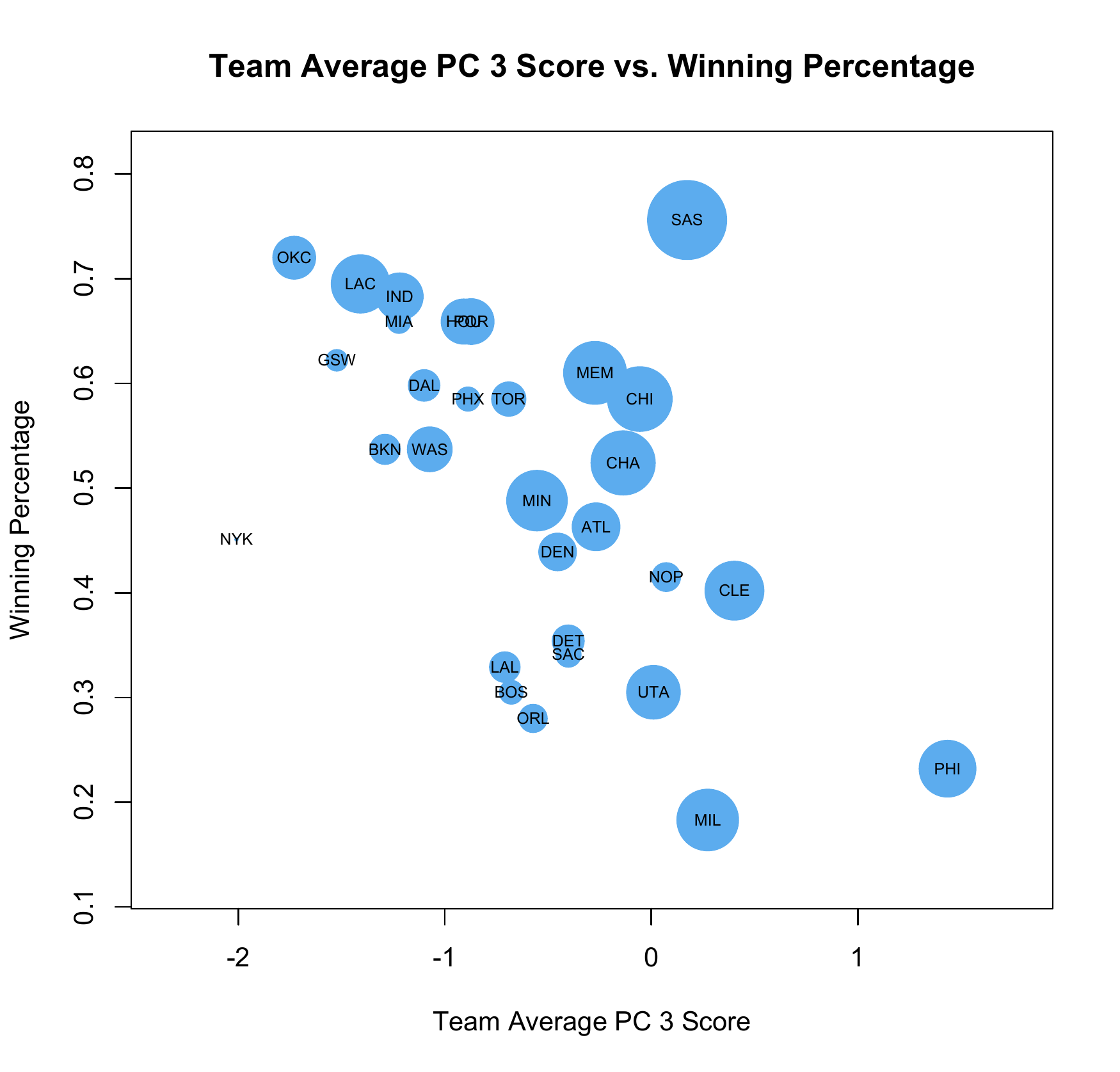}
\caption{\label{fig:pcscoreteam3} Team average PC 3 scores by regular season winning percentage.  Point size determined by regression-weighted sum of PC 2 and PC 4 scores (color online).}
\end{figure}

These results are better illustrated in Figure \ref{fig:pcscoreteam3} where the regression-weighted sum of PC 2 and PC 4 scores ($0.17*$[PC 2 Score] + $0.09*$[PC 4 Score]) determine the size of the points.  Low average PC 3 scores (i.e. high average scoring and rebounding efficiency) have the strongest influence on winning percentages.  However, controlling for PC 3 scores, teams with higher regression-weighted PC 2 and PC 4 scores (i.e. higher average passing, touches, and assisting) generally hold higher winning percentages.  This helps account for the success of the Spurs, who distributed the scoring responsibilities more evenly across the team, resulting in lower average scoring efficiency and higher passing on average (9 of the 13 Spurs players were above the average 58 passes per 48 minutes).

\section{Statistical Diversity Index (SDI)}
\label{sec:SDI}
Another way the PC scores can be used to compare player statistical profiles is to combine them to produce one measure of how different one player's statistical profile is from another.  Here a simple calculation based on the sum of squared difference between the two players' four PC scores is proposed.  

\subsection{Calculation}
For any two players, player $i$ and player $j$, where $t_{k(i)}$ represents the $k$th principal component score for player i, the Statistical Diversity Index (SDI) can be calculated as
\begin{equation}
SDI_{ij} = \sum_{k=1}^4 (t_{k(i)} - t_{k(j)})^2
\end{equation}

\subsection{Personnel Management}

A large SDI for two players indicates that their statistical profiles are very different across the four principal components defined above.  Using this measure, we can develop lists of players that have statistical profiles most similar to certain players which has many applications in personnel management.  

\subsubsection{Case Study 1: Tony Parker}
For coaches and general managers that have a certain player in mind they would like to add to their team, this measure can produce a list of players with the most similar statistical profiles who may also be good candidates to consider and might come at a lower price tag.  For example, Table \ref{tab:parker} lists the five players with the lowest SDI when compared with Tony Parker, meaning they have similar PC scores to Tony Parker, along with their 2013-2014 season salary.

\begin{table}[h]
\centering
\begin{tabular}{|l|c|c|}
\hline
Player & SDI & Salary \\\hline 
Jose Juan Barea (MIN) & 0.7 & \$4,687,000\\
Brandon Jennings (DET) & 2.8 & \$7,655,503 \\
Mike Conley (MEM) & 5.5 & \$8,000,001 \\
Ty Lawson (DEN) & 5.9 &\$10,786,517 \\
Jeff Teague (ATL) & 6.1 & \$8,000,000 \\
\hline 
\end{tabular}
\caption{\label{tab:parker}Players with lowest SDI compared with Tony Parker and 2013-2014 salary from \url{http://www.basketball-reference.com/}}
\end{table}

\begin{figure}
\centering
\begin{subfigure}{.5\textwidth}
  \centering
  \includegraphics[width=1\linewidth]{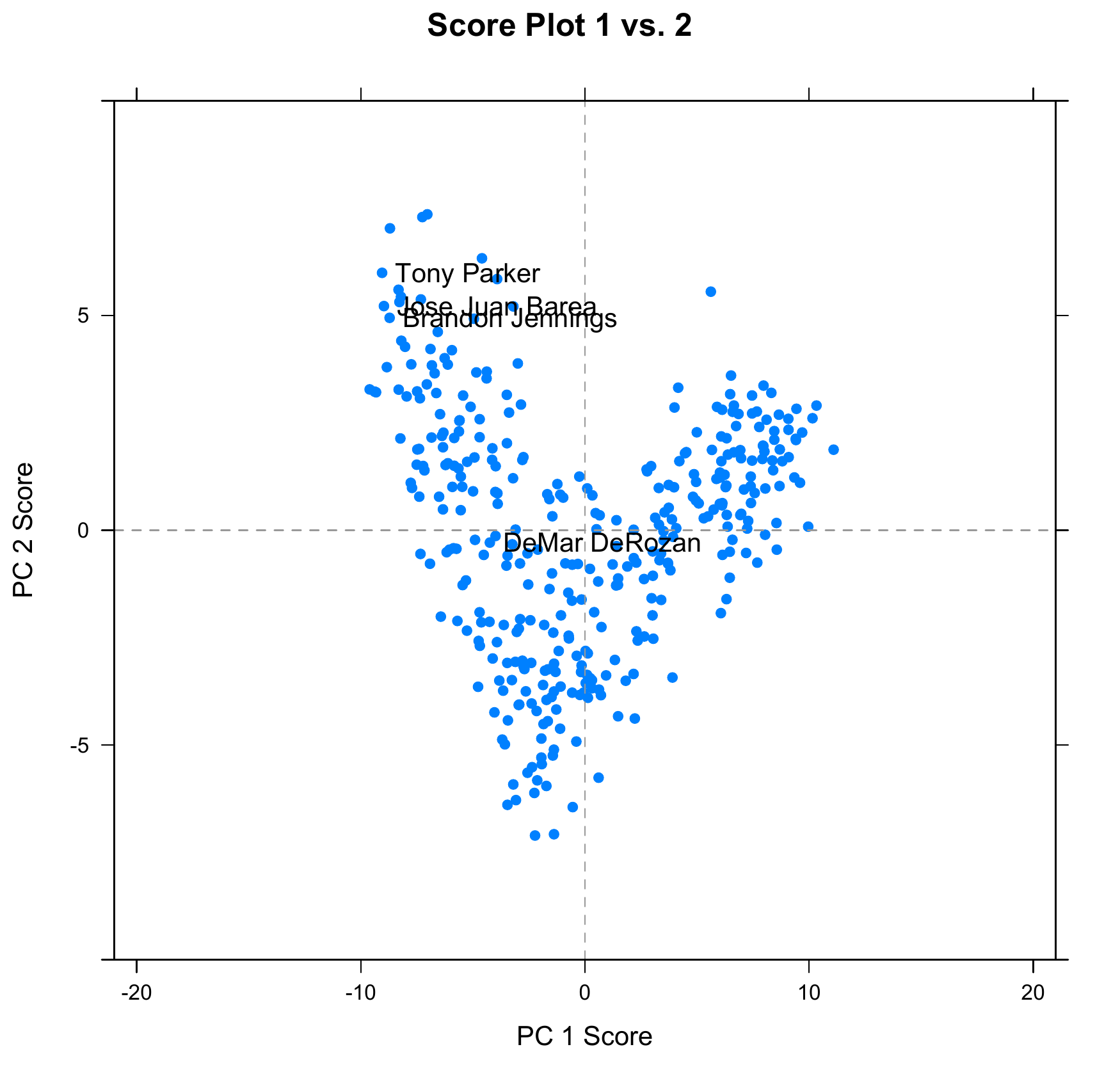}
\end{subfigure}%
\begin{subfigure}{.5\textwidth}
  \centering
  \includegraphics[width=1\linewidth]{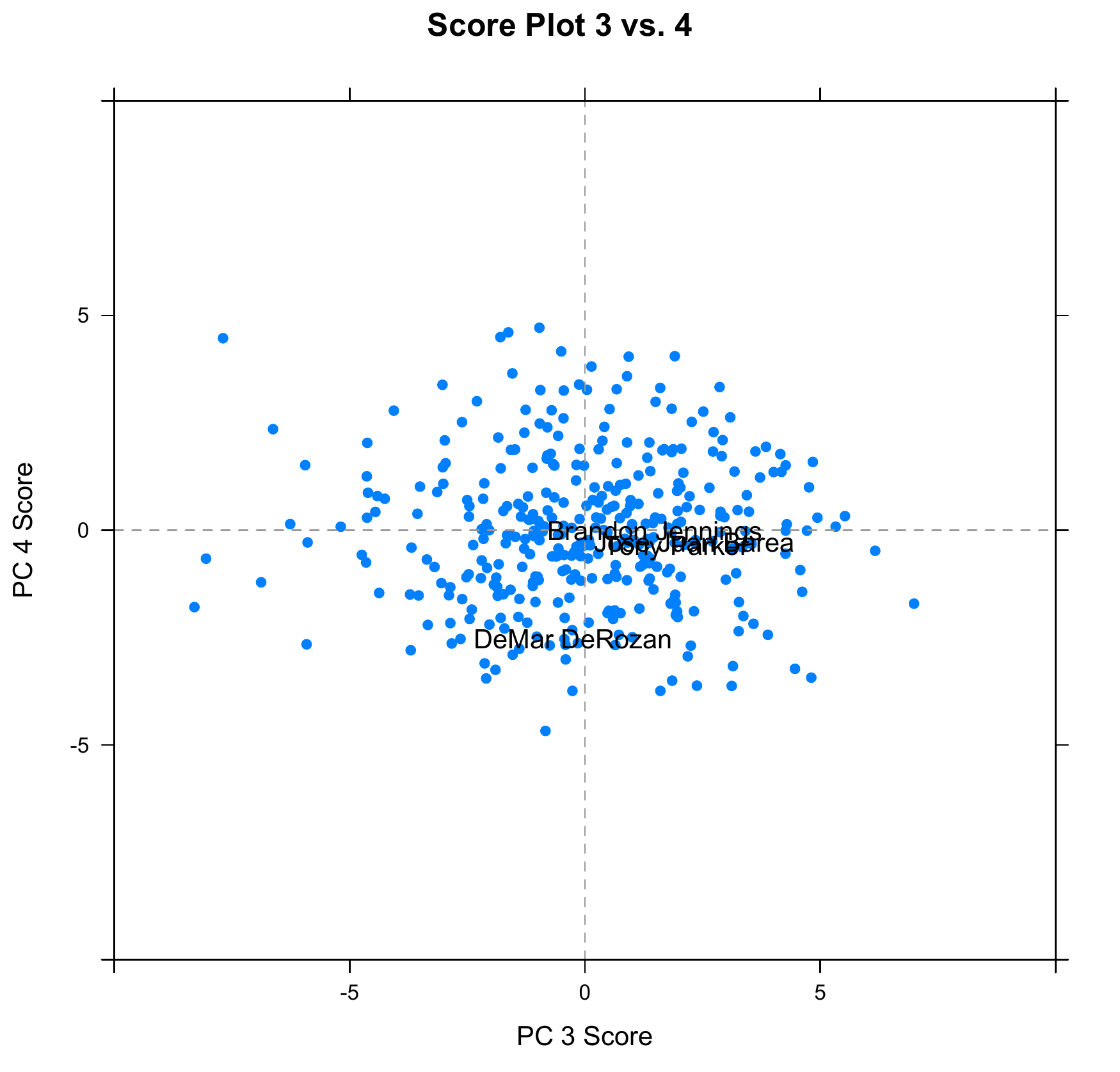}
\end{subfigure}
\centering
\caption{Player PC scores for first four components for Tony Parker comparison (color online).}
\label{fig:parkerscore}
\end{figure}

To better understand the additional value of the player tracking data, PC scores and SDI can be recalculated using only traditional statistics.  This approach identifies DeMar DeRozan as the most similar player to Tony Parker although DeRozan and Parker have an SDI of 75 using player tracking statistics (89 other players have a lower SDI compared to Tony Parker).  To better explore the difference between the comparisons using traditional and player tracking statistics, DeRozan is included in the next set of comparisons along with J.J. Barea and Brandon Jennings from Table \ref{tab:parker}.  See Figure \ref{fig:parkerscore} for a comparison of the PC scores and Table \ref{tab:parkerstats} for selected statistics for these players.\\

 DeRozan and Parker have similar point totals per 48 minutes, but Parker generates more driving points compared to catch and shoot points while DeRozan is more balanced.  Additionally, DeRozan doesn't match the others in terms of assist categories, touches, and passes.  This helps explain why DeRozan's PC 2 score is much smaller than the others as catch and shoot offense and lack of passing and assists contribute to negative PC 2 scores.  Shot type, touches, and passes are key aspects of the player tracking statistics that add value by improving player comparisons beyond simple number of points and assists.\\\\
With player tracking statistics, J.J. Barea rises as a less expensive option to Parker's \$12.5M salary who had the most similar statistical performance to Parker in the 2013-2014 season as measured by SDI.  This can be seen in the similarities among PC scores and the selected statistics and shows how SDI can provide a quick method for identifying similar players for further detailed comparisons.\\

\begin{table}[h]
\begin{tabular}{|l|c|c|c|c|}
\hline
Statistic &Parker & Barea & Jennings & DeRozan    \\
\hline
Points & 27.1 & 21.5 & 21.7 & 28.5  \\
Catch and Shoot Points & 2.5 & 3.0 &3.0 & 5.8 \\
Driving Points & 10.3 & 8&4.2 & 6.1  \\
Assist Opportunities & 19.4 & 20.5&21 & 10.4 \\
Assist Points Created &21.9&22.5&23.7&12.6\\
Touches& 123 & 121 & 113 & 76 \\
Passes & 90 & 89 & 84 & 43\\
\hline
\end{tabular}
\vspace{0.25cm}
\caption{\label{tab:parkerstats}Selected statistics for comparison (per 48 minutes unless stated otherwise).}
\end{table}

\subsubsection{Case Study 2: Anthony Morrow}

Another situation where SDI can be useful is in finding suitable replacements for players who may be considering free agency.  Finding candidates who can play a similar role in the organization can be difficult, but SDI can help identify candidates who may be more prepared to step into a specific role that needs to be filled.  For example, Anthony Morrow left the Pelicans through free agency after the 2013-2014 season to join the Thunder who offered Morrow \$3.2M for 2014-2015 compared to the \$1.15M under his contract with the Pelicans\cite{morrow}.  Using SDI, the Pelicans can find players similarly suited to replace Morrow's shooting ability at 54\% effective shooting percentage and 60\% catch and shoot effective shooting percentage (see Table \ref{tab:morrow}).\\

\begin{table}[h]
\centering
\begin{tabular}{|l|c|c|}
\hline
Player & SDI & Salary \\\hline 
Klay Thompson (GSW) & 2.3 & \$2,317,920\\
CJ Miles (CLE) & 2.9 & \$2,225,000 \\
Tim Hardaway Jr. (NYK) & 3.0 & \$1,196,760 \\
Terrence Ross (TOR) & 3.8 &\$2,678,640 \\
Martell Webster (WAS) & 4.1 & \$5,150,000 \\
\hline 
\end{tabular}
\caption{\label{tab:morrow}Players with lowest SDI compared with Anthony Morrow and 2013-2014 salary from \url{http://www.basketball-reference.com/}}
\end{table}

\begin{figure}
\centering
\begin{subfigure}{.5\textwidth}
  \centering
  \includegraphics[width=.9\linewidth]{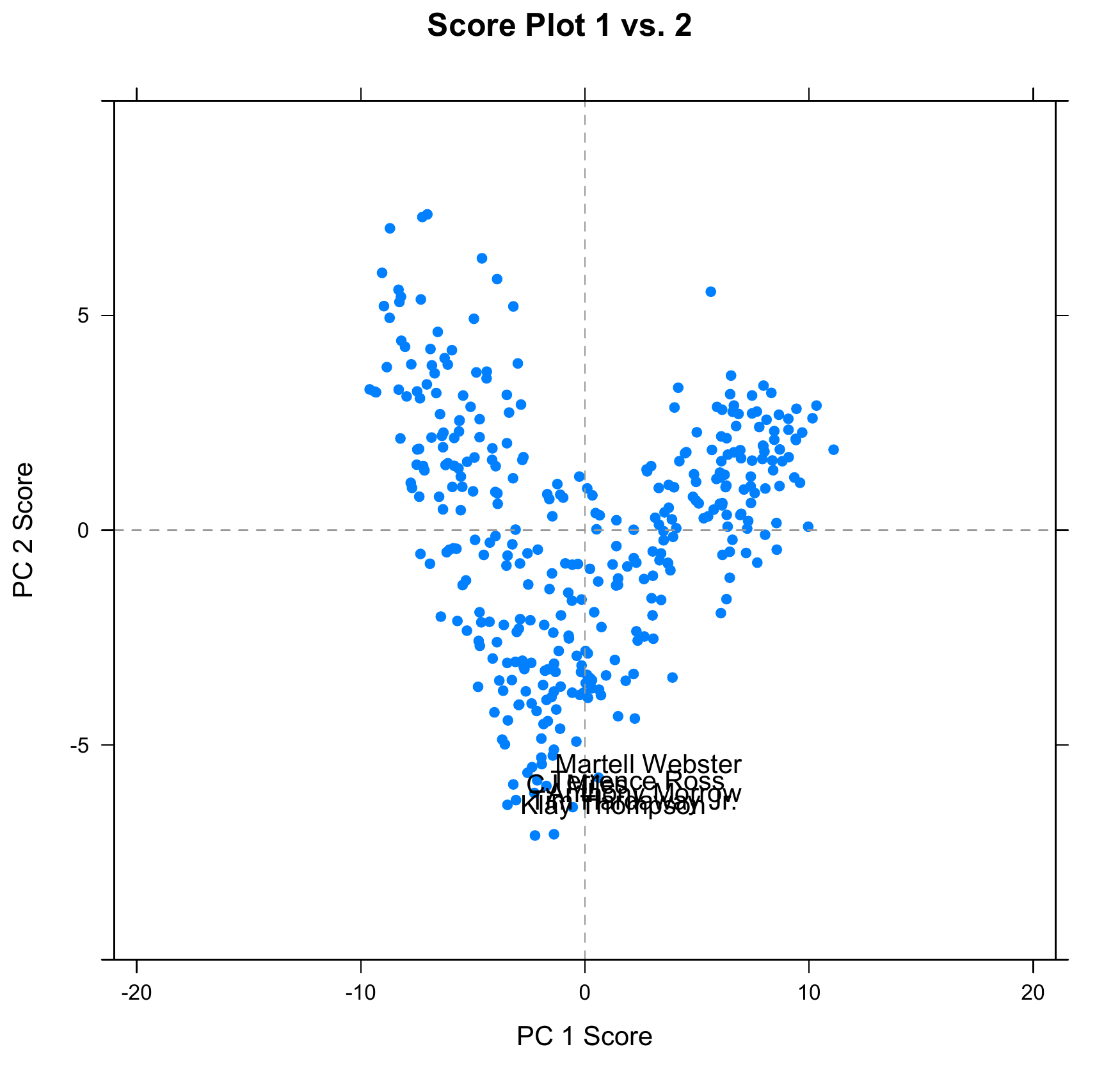}
\end{subfigure}%
\begin{subfigure}{.5\textwidth}
  \centering
  \includegraphics[width=.9\linewidth]{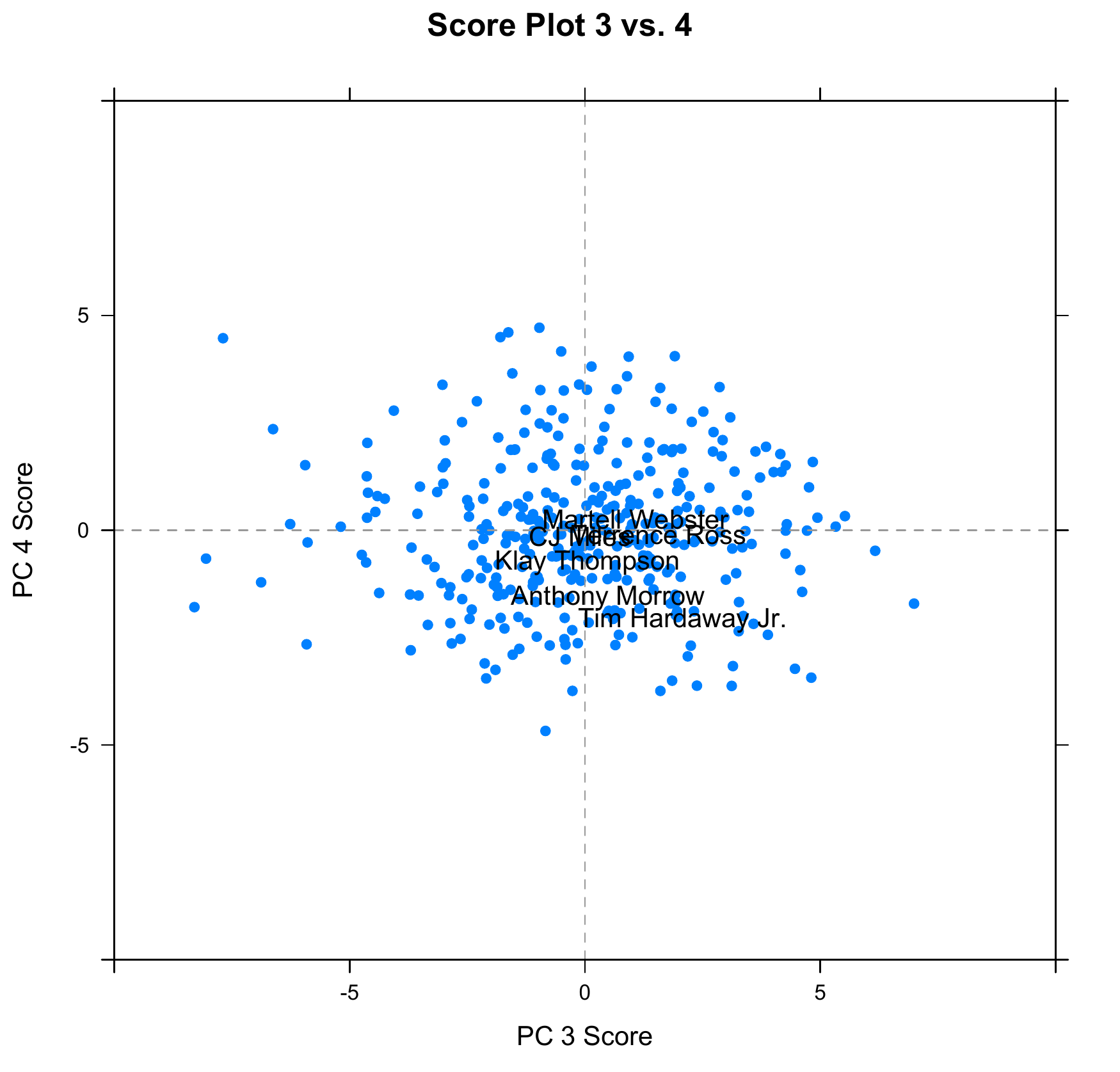}
\end{subfigure}
\centering
\caption{Player PC scores for first four components for Anthony Morrow comparison (color online).}
\label{fig:morrowscore}
\end{figure}

Figure \ref{fig:morrowscore} shows that these players are very similar in terms of the first two PC scores with slight differences in PC 3 and 4.  Based on the interpretation of the PC dimensions previously covered, these players generally produce offense through outside catch and shoot shots (negative PC 1 and 2 scores), but they vary more in shooting efficiency and passing (PC 3 and 4 scores).  See Table \ref{tab:morrowstats}.

\begin{table}[h]
\begin{tabular}{|l|c|c|c|c|c|c|}
\hline
Statistic &Morrow & Thompson & Miles & Hardaway Jr. & Ross & Webster   \\
\hline
Catch and Shoot Points & 9.7 & 12.3 &11.4 & 9.3 & 10.4 & 9.7 \\
Points per Touch & 0.45 & 0.48 & 0.40 & 0.37 & 0.38 & 0.28 \\
Passes & 26.3 & 25.1 & 36.9 & 35.3 & 31.6 & 43.5\\
\hline
\end{tabular}
\vspace{0.25cm}
\caption{\label{tab:morrowstats}Selected statistics for comparison (per 48 minutes unless stated otherwise).}
\end{table}

In terms of catch and shoot points per 48 minutes, all of these players including Morrow are in the top 20 with the exception of Hardaway Jr. at 34th.  However you can better see the differences among the players in terms of points per touch and passes which impact PC 3 and 4 scores.  As SDI increases, the similarity between the players and Morrow's points per touch and passes begin to break down.  However, the Pelicans could use SDI not only to identify suitable replacements to pursue in the offseason but also to estimate the salary of comparable players for use in salary negotiations.

\section{Discussion and Further Remarks}
\label{sec:conclusion}

This article explores the utility of the newly available NBA player tracking data in evaluating and comparing player and team abilities and playing styles through their statistical profiles.  Using PCA, we identify and interpret four principal components that capture over 68\% of the variance in the NBA player tracking data set and compare players along these new dimensions.  A simple measure of comparison between two players or teams based on principal component scores, SDI, is also introduced.  This framework is scalable as it can incorporate existing and new statistics that will emerge to reconstruct the principal component dimensions and SDI for improved comparisons. \\

SDI and the principal component scores can be used by head coaches and general managers to evaluate team performance and personnel needs and also for quickly identifying players with similar statistical profiles to a certain player of interest for use in numerous personnel management applications (e.g. salary negotiations, free agency acquisitions, replacement options for key role players, etc.).  This approach is advantageous as it allows for use of the entirety of the available data for finding suitable comparisons across the NBA quickly along with principal component scores to help understand why players are deemed similar statistically.  This can serve as a starting point for more detailed comparisons by considerably narrowing down the number of players under consideration.\\

This work could be extended by incorporating new data from the 2014-2015 season to see if principal component dimensions and player principal component scores change significantly from one season to another. Also this would provide data for players who didn't see much playing time in the 2013-2014 season.  Additionally, player tracking data at the game level could greatly extend this analysis by tracking how players' statistical profiles change throughout the course of the season.

\clearpage
\appendix
\section*{Appendix}
\addcontentsline{toc}{section}{Appendices}
\renewcommand{\thesubsection}{\Alph{subsection}}

\subsection{Brief Introduction to Principal Components Analysis}
\label{app:pcaintro}
The method can be formulated\cite{pca} as an orthogonal linear transformation of the data into a new coordinate system such that the first direction (first principal component) contains the greatest variance in the data, the second direction (second principal component) contains the second greatest variance, etc.  Consider a data matrix $\mathbf{X}_{nxp}$ whose $n$ rows represent observations each with $p$ different variables of interest.  We define a set of $px1$ loading vectors, $\mathbf{w}_{(k)}$, $k=1,\ldots,p$ that map each row of observations in  $\mathbf{X}_{nxp}$, call it $\mathbf{x}_{(i)}$, $i=1,\ldots,n$ to a new vector of principal component scores, call it $\mathbf{t}_{(i)}$ such that $\mathbf{t}_{k(i)} = \mathbf{x}_{(i)}*\mathbf{w}_{(k)}$.  We can find the loading vector for the first principal component as $\mathbf{w}_{(1)}$ such that the variance of the corresponding principal component scores $\mathbf{t}_{1(i)}$ is maximized and $\mathbf{w}_{(1)}$ is of unit length, which can be expressed as:
$$\mathbf{w}_{(1)} = \mathrm{arg max}\frac{\mathbf{w}^T\mathbf{X}^T\mathbf{X}\mathbf{w}}{\mathbf{w}^T\mathbf{w}}$$ 
The remaining principal component loading vectors can be found in a similar fashion.  To find the $k$th component, subtract the first k-1 components from $\mathbf{X}$:
$$\hat{\mathbf{X}}_k = \mathbf{X} - \sum_{s=1}^{k-1}\mathbf{X}\mathbf{w}_{(s)}\mathbf{w}_{(s)}^T$$
Then use the same variance maximization method on the new matrix $\hat{\mathbf{X}}_k$ to find the $k$th component:
$$\mathbf{w}_{(k)} = \mathrm{arg max}\frac{\mathbf{w}^T\hat{\mathbf{X}}_k^T\hat{\mathbf{X}}_k\mathbf{w}}{\mathbf{w}^T\mathbf{w}}$$ 
Principal component analysis can also be viewed as the eigenvalue decomposition of the covariance matrix where the eigenvectors are the principal components as stated above.  There are numerous resources available for explaining the methodology in more detail, but a basic understanding as outlined above should suffice for this article. 
\newpage
\subsection{Principal Component Loading Vectors} 
\label{app:loadings}
Variable loadings for coefficients greater than 0.1 in absolute value.  Statistics are per 48 minutes unless otherwise stated.

\begin{table}[h]
\centering
\begin{tabular}{l|c}
Statistic & Loading\\\hline
Contested Rebounds & 0.178\\
Rebounding Opportunities & 0.176\\
Offensive Rebounds & 0.173\\
Total Rebounds & 0.172\\
Contested Offensive Rebounds & 0.171\\
Opponent Field Goal Attempts at the Rim & 0.17\\
Contested Defensive Rebounds & 0.17\\
Uncontested Rebounding Percentage & 0.169\\
Opponent Field Goal Makes at the Rim & 0.169\\
Offensive Rebounding Opportunities & 0.169\\
Defensive Rebounding Opportunities & 0.166\\
Close Points & 0.161\\
Close Attempts & 0.159\\
Close Touches & 0.158\\
Defensive Rebounds & 0.156\\
Uncontested Rebounds & 0.154\\
Uncontested Offensive Rebounds & 0.153\\
Uncontested Defensive Rebounding Percentage & 0.153\\
Uncontested Offensive Rebounding Percentage & 0.145\\
Blocks & 0.14\\
Uncontested Defensive Rebounds & 0.14\\
Points per Half Court Touch & 0.121\\
Elbow Touches & 0.114\\
Catch and Shoot Effective Field Goal Percentage & -0.104\\
Catch and Shoot Attempts & -0.105\\
Catch and Shoot Makes & -0.106\\
Pull Up 3-Point Percentage & -0.116\\
Secondary Assists & -0.117\\
Free Throw Assists & -0.127\\
Pull Up 3-Point Makes & -0.128\\
Driving Points & -0.132\\
Driving Attempts & -0.137\\
Assists & -0.137\\
Pull Up 3-Point Attempts & -0.137\\
Points Created From Assists & -0.137\\
Catch and Shoot 3-Point Percentage & -0.139\\
Assist Opportunities & -0.14\\
Time of Possession & -0.142\\
Pull Up Makes & -0.143\\
Pull Up Points & -0.145\\
Team Driving Points & -0.146\\
Drives & -0.146\\
Pull Up Attempts & -0.15\\
Front Court Touches & -0.15
\end{tabular}
\caption{\label{tab:pc1}Significant statistic loadings for Principal Component 1}
\end{table}

\begin{table}[h]
\centering
\begin{tabular}{l|c}
Statistic & Loading\\\hline
Touches & 0.246\\
Passes & 0.237\\
Assist Opportunities & 0.208\\
Assists & 0.206\\
Points Created From Assists & 0.206\\
Time of Possession & 0.203\\
Secondary Assists & 0.19\\
Free Throw Assists & 0.183\\
Front Court Touches & 0.18\\
Team Driving Points & 0.147\\
Drives & 0.146\\
Driving Attempts & 0.125\\
Driving Points & 0.125\\
Points per Half Court Touch & -0.114\\
Catch and Shoot 3-Point Percentage & -0.115\\
Points per Touch & -0.159\\
Catch and Shoot 3-Point Makes & -0.234\\
Catch and Shoot 3-Point Attempts & -0.235\\
Catch and Shoot Makes & -0.242\\
Catch and Shoot Attempts & -0.251\\
Catch and Shoot Points & -0.254\\
\end{tabular}
\caption{\label{tab:pc2}Significant statistic loadings for Principal Component 2}
\end{table}

\begin{table}[h]
\centering
\begin{tabular}{l|c}
Statistic & Loading\\\hline
Average Defensive Speed & 0.306\\
Distance & 0.3\\
Average Speed & 0.3\\
Average Offensive Speed & 0.226\\
Opponent Points at the Rim & 0.202\\
Close Shooting Percentage & 0.156\\
Pull Up 3-Point Makes & -0.108\\
Catch and Shoot Effective Shooting Percentage & -0.113\\
Effective Shooting Percentage & -0.113\\
Catch and Shoot Attempts & -0.113\\
Defensive Rebounds & -0.114\\
Offensive Rebounding Percentage & -0.119\\
Elbow Touches & -0.122\\
Catch and Shoot Points & -0.124\\
Uncontested Defensive Rebounds & -0.124\\
Pull Up Attempts & -0.138\\
Catch and Shoot Makes & -0.146\\
Catch and Shoot Percentage & -0.148\\
Pull Up Points & -0.151\\
Pull Up Makes & -0.153\\
Points per Half Court Touch & -0.173\\
Defensive Rebounding Percentage & -0.212\\
Points per Touch & -0.232\\
Rebounding Percentage & -0.235\\
Points & -0.305\\
\end{tabular}
\caption{\label{tab:pc3}Significant statistic loadings for Principal Component 3}
\end{table}

\begin{table}[h]
\centering
\begin{tabular}{l|c}
Statistic & Loading\\\hline
Passes & 0.311\\
Touches & 0.25\\
Catch and Shoot Makes & 0.248\\
Catch and Shoot Percentage & 0.237\\
Catch and Shoot Points & 0.236\\
Catch and Shoot Effective Shooting Percentage & 0.224\\
Catch and Shoot Attempts & 0.212\\
Average Offensive Speed & 0.209\\
Catch and Shoot 3-Point Makes & 0.166\\
Distance & 0.151\\
Secondary Assists & 0.15\\
Average Speed & 0.146\\
Defensive Rebounding Opportunities & 0.142\\
Catch and Shoot 3-Point Attempts & 0.141\\
Uncontested Defensive Rebounds & 0.139\\
Defensive Rebounds & 0.137\\
Uncontested Rebounds & 0.117\\
Contested Defensive Rebounds & 0.115\\
Opponent Makes at the Rim & 0.114\\
Elbow Touches & 0.102\\
Opponent Attempts at the Rim & 0.101\\
Drives & -0.13\\
Team Driving Points & -0.137\\
Points per Half Court Touch & -0.151\\
Driving Attempts & -0.167\\
Drives & -0.193\\
Points per Touch & -0.264\\
\end{tabular}
\caption{\label{tab:pc4}Significant statistic loadings for Principal Component 4}
\end{table}
\end{document}